\documentclass[sigconf]{acmart}
\usepackage[english]{babel}

\usepackage{amsmath}
\usepackage{mathrsfs}
\usepackage{algorithm,algpseudocode}
\usepackage{color,xcolor}
\usepackage{algpseudocode}
\usepackage{amsmath}
\usepackage{balance}
\usepackage{graphics}
\usepackage{epsfig}
\usepackage{balance}
\usepackage{threeparttable}
\usepackage{enumitem}
\usepackage{bm}
\usepackage{verbatim}
\usepackage{soul}
\setstcolor{red}
\usepackage{multirow}
\usepackage{makecell}
\usepackage{xspace}
\usepackage{appendix}
\usepackage{booktabs}
\usepackage{multirow}
\usepackage{adjustbox}
\usepackage[justification=justified, skip=5pt]{caption}

\newtheorem{myDef}{Definition}

\newcommand{\eg}{\emph{e.g.,}\xspace}
\newcommand{\ie}{\emph{i.e.,}\xspace}
\newcommand{\wrt}{\emph{w.r.t.}\xspace}
\newcommand{\model}{PRS}
\newcommand{\RM}{PMatch}
\newcommand{\RR}{PRank}
\newcommand{\FLSA}{FPSA}
\newcommand{\DCWN}{DPWN}

\AtBeginDocument{%
  \providecommand\BibTeX{{%
    \normalfont B\kern-0.5em{\scshape i\kern-0.25em b}\kern-0.8em\TeX}}}

\begin{document}
\title{Revisit Recommender System in the Permutation Prospective}

\author{Yufei Feng, Yu Gong, Fei Sun, Junfeng Ge, Wenwu Ou}

\affiliation{%
\institution{Alibaba Group, Hangzhou, China} 
}
\email{
fyf649435349@gmail.com, {gongyu.gy, ofey.sf, beili.gjf}@alibaba-inc.com, santong.oww@taobao.com
}

\begin{abstract}
Recommender systems (RS) work effective at alleviating information overload and matching user interests in various web-scale applications. Most RS retrieve the user's favorite candidates and then rank them by the rating scores in the greedy manner.
In the permutation prospective, however, current RS come to reveal the following two limitations: 
1) They neglect addressing the permutation-variant influence within the recommended results;
2) Permutation consideration extends the latent solution space exponentially, and current RS lack the ability to evaluate the permutations. 
Both drive RS away from the permutation-optimal recommended results and better user experience.

To approximate the permutation-optimal recommended results effectively and efficiently, we propose a novel permutation-wise framework {\model} in the re-ranking stage of RS, which consists of Permutation-Matching ({\RM}) and Permutation-Ranking ({\RR}) stages successively. 
Specifically, the {\RM} stage is designed to obtain the candidate list set, where we propose the {\FLSA} algorithm to generate multiple candidate lists via the permutation-wise and goal-oriented beam search algorithm.
Afterwards, for the candidate list set, the {\RR} stage provides a unified permutation-wise ranking criterion named LR metric, which is calculated by the rating scores of elaborately designed permutation-wise model {\DCWN}. 
Finally, the list with the highest LR score is recommended to the user. 
Empirical results show that {\model} consistently and significantly outperforms state-of-the-art methods. Moreover, {\model} has achieved a performance improvement of 11.0\% on PV metric and 8.7\% on IPV metric after the successful deployment in one popular recommendation scenario of Taobao application. 
\end{abstract}




\maketitle

\section{Introduction}

\begin{figure}
    \centering
    \includegraphics[scale=0.65]{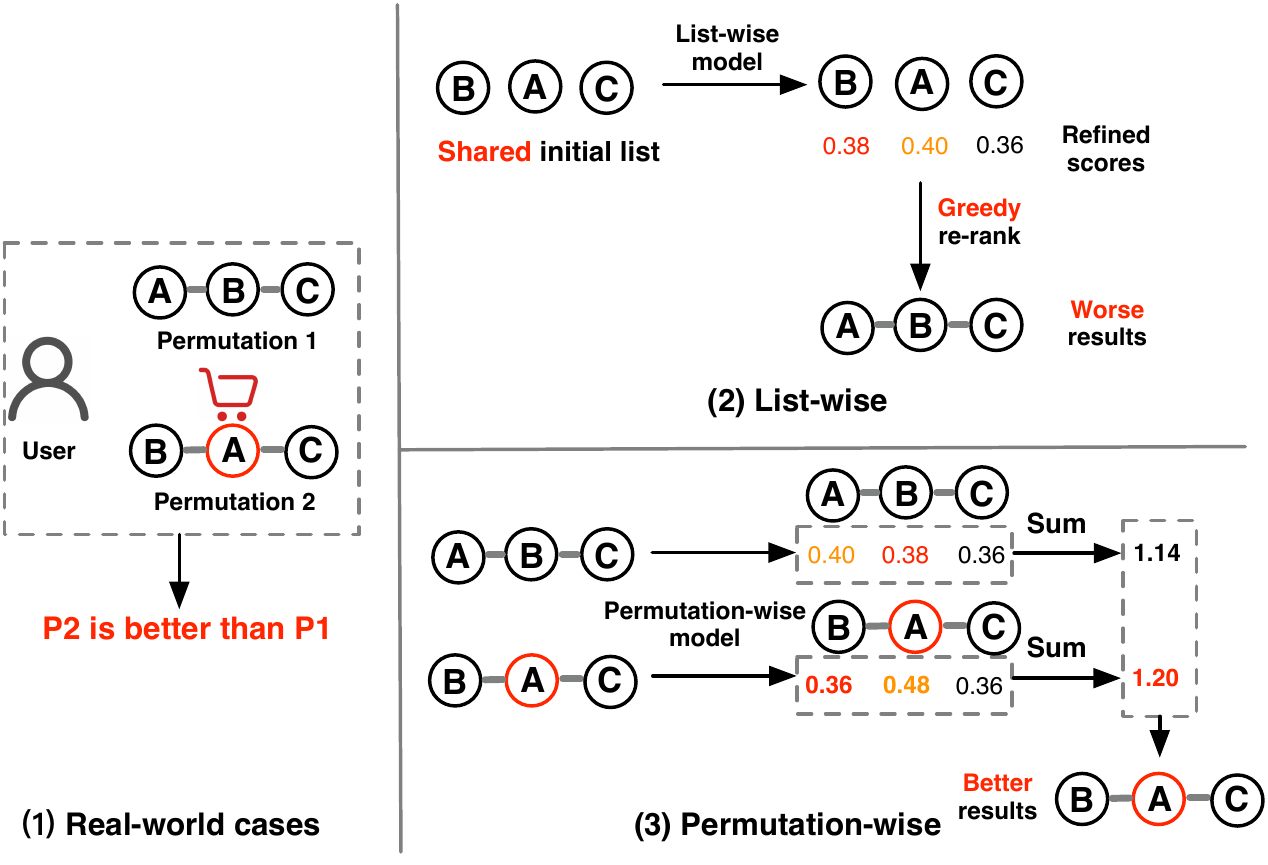}
    \caption{(1) shows two real-world cases. (2) and (3) present the comparison of list-wise and permutation-wise methods when meeting the two cases.}
    \label{fig:intro_0}
\end{figure}

Recommender systems (RS) have been widely deployed in various web-scale applications, including e-commerce~\citeN{Sarwar:ItemCF, Paul:YoutubeNet, Zhou:DIN, Feng:DSIN}, social media~\citeN{Paul:YoutubeNet, gomez2015netflix} and news~\cite{das2007google, liu2010personalized, Li:SCENE}. 
Typically, web-scale RS usually consist of three stages successively, i.e., matching, ranking, and re-ranking.
With the continuous advances in the recommendation technologies, various methods have been proposed in each stage to improve user experience and recommendation performance. 
Overall, most methods in RS contribute to retrieve the user's favorite items from tremendous candidates, followed by the greedy strategy based on the rating scores of given user-item pairs.

Different from point-wise methods in the matching and ranking stages, various list-wise methods~\citeN{Ai:DLCM, Pei:PRM, Zhuang:MIDNN, pang2020setrank} have been proposed and deployed in the re-ranking stage. The widely adopted pipeline of existing list-wise methods usually contain three key components: 1) \textit{Ranking}: the initial list is generated according to scores of the basic ranking model; 2) \textit{Refining}: the list-wise feature distribution of the initial list is extracted by a well-designed module (\eg LSTM~\cite{Ai:DLCM} and self-attention~\citeN{pang2020setrank, Pei:PRM}) to refine the rating scores; 3) \textit{Re-ranking}: candidate items are re-ranked by the refined list-wise rating scores in the greedy manner. 
Overall, existing list-wise methods achieve improvements in recommendations mainly by modeling list-wise feature distribution to focus on refining the item's rating scores.

In the permutation prospective, however, the popular re-ranking pipeline cannot assist current RS to reach permutation-optimal due to neglecting permutation-variant influence within the recommended results. 
As shown in Fig.~\ref{fig:intro_0} (1), here are two recommended results collected from the real-world dataset. Surprisingly, even if they are composed of the \textit{same} items and exhibited to the \textit{same} user, the user responses to the permutation $2$ rather than $1$. One possible reason is that placing the more expensive item $B$ ahead can stimulate the user's desire to buy the cheaper item $A$. We refer such influence on the user's making decisions caused by the change of permutations as \textit{permutation-variant influence} within the recommended results. By addressing such permutation-variant influence, the ideal method should recommend permutation $2$ to the user, which is the \textit{permutation-optimal} result in the permutation prospective. While with the supervised training of item $A$ in the permutation $2$, the list-wise models in Fig.~\ref{fig:intro_0} (2) can only acquire the priority of item $A$ when meeting the shared initial list (\eg item $B$, $A$, $C$). Afterwards, the list-wise models will re-rank items in the greedy manner by the refined rating scores (\ie permutation $1$) and miss the better permutation $2$. Ideally, with the permutation-wise methods in Fig.~\ref{fig:intro_0} (3), the predicted interaction probability of item $A$ has been greatly improved by placing item $B$ ahead of $A$ from permutation $1$ to $2$, which mainly contributes to the correct judgment of permutation $2$ better than $1$. Accordingly, fully considering and leveraging such permutation-variant influence contained in lists serve as the key to approximate the permutation-optimal recommended results and better user experience.

Practically, such permutation prospective brings new challenges to current RS, which can mainly be summarized as the following two aspects: 
1) \textbf{Exponential solutions.} Suppose $n$ (\eg $10$) items need to be recommended from the initial list of size $m$ (\eg $100$). Current RS regard it as a retrieval task, that is, deploying list-wise models in the re-ranking stage to search in the O(m) solution space. In the permutation prospective, however, the solution space swells from O(m) to O($\mathrm{A}_m^n$) (about O($100^{10}$)). 
Considering the permutation-variant influence within the final item lists, the user will response differently to each list. In fact, only one list will be finally recommended to the user, thus how to reach the permutation-optimal list effectively and efficiently brings new challenges to current RS; 
2) \textbf{Permutation-wise evaluation.} Most methods applied in current RS attempt to point-wisely predict the user-item interaction probability (\eg click-through rate and conversion rate). As mentioned above, it is necessary to provide a unified permutation-wise ranking criterion for selecting the permutation-optimal one from massive qualified lists, which has not been involved in the existing efforts. 

\begin{figure}
    \centering
    \includegraphics[scale=0.65]{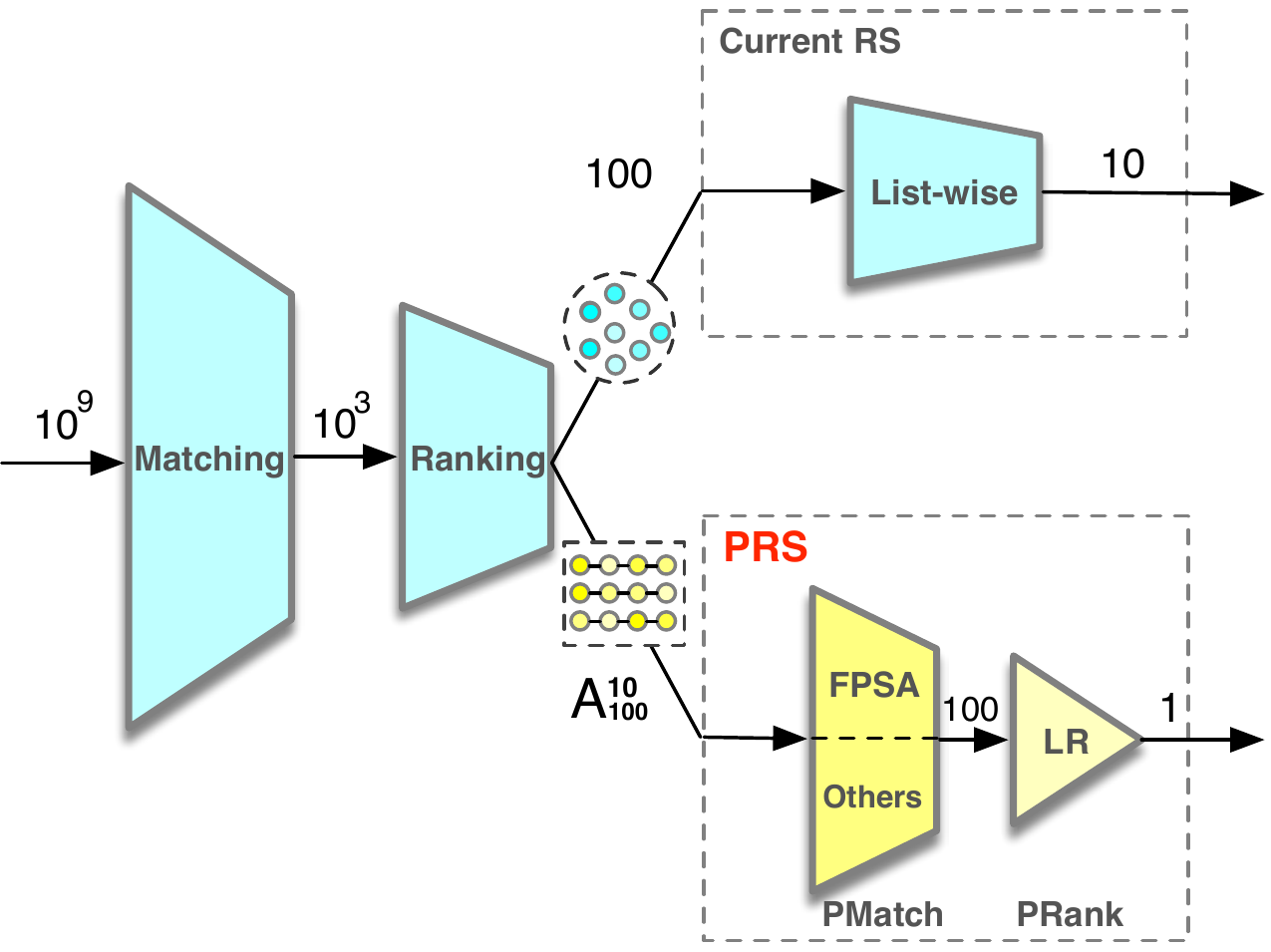}
    \caption{Difference of architecture between current RS and {\model} in the re-ranking stage.} 
    \label{fig:framework}
\end{figure}

In this work, as shown in Fig.~\ref{fig:framework}, we promote the list-wise methods to a novel permutation-wise framework named {\model} (\textbf{P}ermutation \textbf{R}etrieve \textbf{S}ystem) in the re-ranking stage of RS. Specifically, {\model} consists of {\RM} (\textbf{P}ermutation-\textbf{M}atching) and {\RR} (\textbf{P}ermutation-\textbf{R}anking)  stages successively.
The {\RM} stage focuses on generating multiple candidate lists in an efficient and parallel way, with the aim of considering permutation-variant influence and alleviating the problem of exponential solutions. 
Here we propose {\FLSA} (\textbf{F}ast \textbf{P}ermutation \textbf{S}earching \textbf{A}lgorithm), a permutation-wise and goal-oriented beam search algorithm, to generate candidate lists in an efficient way.
Afterwards, the {\RR} stage provides a unified ranking criterion for the challenge of permutation-wise evaluation. We equip the proposed model {\DCWN} (\textbf{D}eep \textbf{P}ermutation-\textbf{W}ise \textbf{N}etwork) with Bi-LSTM, with which permutation-variant influence can be fully addressed. Moreover, we propose the LR (\textbf{L}ist \textbf{R}eward) metric to rank the candidate lists, which is calculated by adding the {\DCWN}'s rating scores of each item within the candidate list. Finally, the list with the highest LR score will be recommended to the user. 

To demonstrate the effectiveness of {\model}, we perform a series of experiments on a public dataset from Alimama and a proprietary dataset from Taobao application. Experimental results demonstrate that {\model} consistently and significantly outperforms various state-of-the-art methods. Moreover, {\model} has been successfully deployed in one popular recommendation scenario of Taobao application and gained a performance improvement of 11.0\% on the PV (Page View) metric and 8.7\% on the IPV (Item Page View) metric.

\section{Related Work}
In this section, we review the most related studies in the re-ranking stage in both academic and industrial recommender systems. Typically, the re-ranking stage serves as the last stage after the matching and ranking stages, where an initial ranking list obtained from previous stages is re-ranked into the final item list (\ie recommended results) to better satisfy user demands. Roughly speaking, existing re-ranking methods mainly fall into three categories: point-wise, pair-wise and list-wise methods. 
Point-wise methods~\citeN{Paul:YoutubeNet, Cheng:WDL, Guo:DeepFM} regard the recommendation task as a binary classification problem and globally learn a scoring function for a given user-item pair with manually feature engineering. Though with continuous achievements, these methods neglect considering the comparison and mutual influence between items in the input ranking list. 
To solve this problem, pair-wise and list-wise models are gradually paid more and more attention in current RS. Pair-wise models work by considering the semantic distance of an item pair a time. 
RankSVM~\cite{Joachims:RankSVM}, GBRank~\cite{Zheng:GBRank} and RankNet~\cite{Burges:RankNet} apply SVM, GBT and DNN with techinically designed pair-wise loss functions to compare any item pair in the input ranking list, respectively. However, pair-wise methods neglect the local information in the list and increase the model complexity.
List-wise models are proposed to capture the interior correlations among items in the list in different ways. 
LambdaMart~\cite{Burges:lambdamart} is a well-known tree-based method with the list-wise loss function.
MIDNN~\cite{Zhuang:MIDNN} works by handmade global list-wise features, while it requires much domain knowledge and decreases its generalization performance. DLCM~\cite{Ai:DLCM}, PRM~\cite{Pei:PRM}, and SetRank~\cite{pang2020setrank} apply GRU, self-attention, and induced self-attention to encode the list-wise information of the input ranking list for better prediction, respectively. 
Though effective, this type of list-wise models does not escape the paradigm of greedy ranking based on rating scores, where the permutation-variant influence within recommended results are not fully addressed.

There are also some other works~\cite{Chen:DPP, Wilhelm:DPPInYoutube, Gelada:DeepMDP, Gogna:BalancingAccAndDiversity} focusing on making the trade-off between relevance and diversity in the re-ranking stage.
Another emerging direction of re-ranking methods is group-wise methods~\cite{ai2019learning}, where the relevance score of a
document is determined jointly by multiple documents in the list. Though effective, group-wise methods may not be appropriate for industrial recommender systems due to its high computation complexity (at least $O(N^2)$).

Note that our proposed {\model} can be easily integrated with  existing works, deploying them in parallel and merging them into the candidate list set in the {\RM} stage.

\section{Preliminary}
In general, a web-scale recommender system (\eg e-commerce and news) is composed of three stages chronologically: matching, ranking and re-ranking. In this paper, we focus on the final re-ranking stage, whose input is the ranking list produced by the previous two stage (\ie matching and ranking). The task of the re-ranking is to elaborately select candidates from the input ranking list and rearrange them into the final item list, followed by the exhibition for users.
Mathematically, with user set $\mathcal{U}$ and item set $\mathcal{I}$, we denote labeled list interaction records as $\mathcal{R} = \{(u, \mathcal{C}, \mathcal{V}, \mathcal{Y}^{\text{CTR}}, \mathcal{Y}^{\text{NEXT}} | u \in \mathcal{U}, \mathcal{V} \subset \mathcal{C} \subset \mathcal{I})\}$.
Here, $\mathcal{C}$ and $\mathcal{V}$ represent the recorded input ranking list with $m$ items for re-ranking stage and the recorded final item list with $n$ items exhibited to user $u$, respectively. Intuitively, $n \leq m$. $y^{\text{CTR}}_{t} \in \mathcal{Y}^{\text{CTR}}$ is the implicit feedback of user $u$ \wrt $t$-th item $v_t \in \mathcal{V}$, where $y^{\text{CTR}}_{t} = 1$ when interaction (\eg click) is observed, and $y^{\text{CTR}}_{t} = 0$ otherwise. Similarly, $y^{\text{NEXT}}_{t} = 1$ indicates that the user continue browsing the followings after this item, and $y^{\text{NEXT}}_{t} = 0$ otherwise. In the real-world industrial recommender systems, each user $u$ is associated with a user profile $x_u$ consisting of sparse features $x^u_s$ (\eg user id and gender) and dense features $x^u_d$ (\eg age), while each item $i$ is also associated with a item profile $x_i$ consisting of sparse features $x^s_i$ (\eg item id and brand) and dense features $x_i^d$ (\eg price). 

Given the above definitions, we now formulate the re-ranking task to be addressed in this paper:
\begin{myDef}
\textbf{Task Description}. Typically, the purpose of industrial RS is to make users take more browsing (PV, page view) and interactions (IPV, item page view), and the same for the re-ranking task. Given a certain user $u$, involving his/her input ranking list $\mathcal{C}$, the task is to learn a re-ranking strategy $\pi: \mathcal{C} \stackrel{\pi}{\longrightarrow} \mathcal{P}$, which aims to select and rearrange items from $\mathcal{C}$, and subsequently recommend a final item list $\mathcal{P}$. 
\end{myDef}

\section{The Proposed Framework}

\begin{table}
  \caption{Key notations.}
  \label{table:notations}
  \begin{tabular}{ll}
    \toprule
    Notations & Description\\
    \midrule
    $\mathcal{U}$, $\mathcal{I}$ & \makecell[{}{p{5.8cm}}]{the set of users and items, respectively}\\
    \hline
    $\mathcal{C}$, $\mathcal{V}$ & \makecell[{}{p{5.8cm}}]{the recorded input ranking list and labeled final item list, respectively}\\
    \hline
    $\mathcal{Y}^{\text{CTR}}, \mathcal{Y}^{\text{NEXT}}$ & \makecell[{}{p{5.8cm}}]{the implicit feedback whether the user clicks or continues browsing of the final item list, respectively}\\
    \hline
    $m, n$ & \makecell[{}{p{5.8cm}}]{the number of items in the recorded input and final ranking list, respectively}\\
    \hline
    $\pi$, $\mathcal{P}$ & \makecell[{}{p{5.8cm}}]{the learned re-ranking strategy and the generated final ranking list, respectively}\\
    \hline
    $x_s^u (x_s^i), x_d^u (x_s^i)$ & \makecell[{}{p{5.8cm}}]{the sparse and dense feature set for users (items), respectively} \\
    \hline 
    $\alpha, \beta$ & \makecell[{}{p{5.8cm}}]{Fusion coefficient float of $r^{PV}$ and $r^{IPV}$ in {\FLSA}, respectively} \\
    \hline
    $r^{PV}$, $r^{IPV}$, $r^{sum}$ & \makecell[{}{p{5.8cm}}]{the estimated PV, IPV and mixed reward of the list, respectively} \\
    \hline
    $\mathcal{S}$ & \makecell[{}{p{5.8cm}}]{the candidate list set generated by the {\RM} stage} \\
    \bottomrule
  \end{tabular}
\end{table}

In this section, we introduce our proposed permutation-wise framework {\model}, which aims to leverage permutation-variant influence effectively and efficiently for better item rearrangements in the re-ranking stage of RS. 
To address the exponential solutions and permutation-wise evaluation challenges, we transform the re-ranking task into two successive stages: Permutation-Matching ({\RM}) and Permutation-Ranking ({\RR}). 
In the {\RM} stage, multiple efficient methods can be deployed in parallel to consider permutation-variant influence and search for effective candidate item lists from exponential solutions. 
Successively, the {\RR} stage proposes the LR metric as a unified permutation-wise ranking criterion for the candidate item list set, with which the list of the highest LR score will be finally recommended. The key notations we will use throughout the article are summarized in Tab.~\ref{table:notations}.

First of all, we begin with the representations of users and items, which are basic inputs of our proposed framework. Following previous works~\cite{Paul:YoutubeNet, Zhou:DIN}, we parameterize the available profiles into vector representations for users and items. Given a user $u$, associated with sparse features $x_u^s$ and dense features $x_u^d$,  we embed each sparse feature value into $d$-dimensional space. Subsequently, each user $u$ can be represented as $\mathbf{x}_u \in \mathbb{R}^{|x^s_u| \times d + |x^d_u|}$, where $|x^s_u|$ and $|x^d_u|$ denote the size of sparse and dense feature space of user $u$, respectively. 
Similarly, we represent each item $i$ as $\mathbf{x}_i \in \mathbb{R}^{|x^s_i| \times d + |x^d_i|}$. 
Naturally, we represent the recorded input ranking list $\mathcal{C}$ as $\mathcal{C} = [\mathbf{x}_c^1, \dots, \mathbf{x}_c^m]$ and the labeled final item list $\mathcal{V}$ as $\mathcal{V} = [\mathbf{x}_v^1, \dots, \mathbf{x}_v^n]$,
where $m$ and $n$ is the number of items in the input ranking list and final item list, respectively.

In the following sections, we shall zoom into the {\RM} and {\RM} stages of proposed permutation-wise framework {\model}.
For each stage, we split the introductions into two parts: one is offline training (preparation) and the other is online serving (inference). The major difference of the two parts is the $\mathcal{V}, \mathcal{Y}^{\text{CTR}}, \mathcal{Y}^{\text{NEXT}}$ in $\mathcal{R}$ is given in the offline training, while not in the online serving.

\begin{figure}
    \centering
    \includegraphics[scale=1.0]{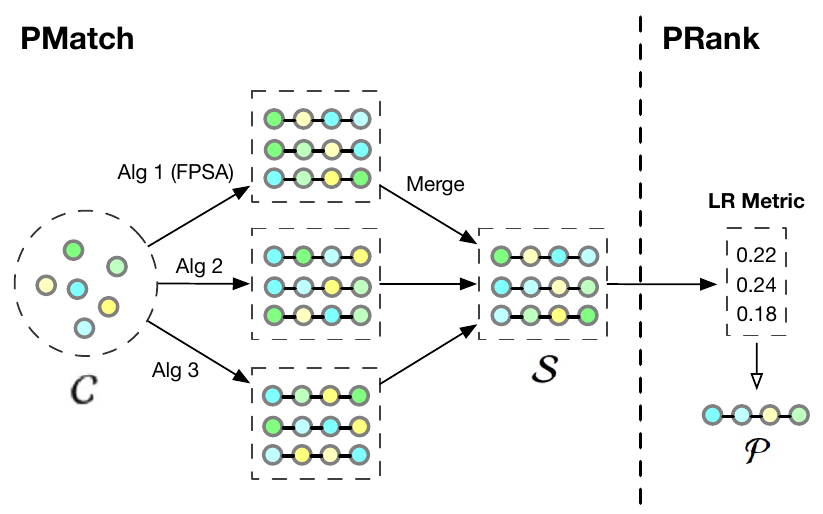}
    \caption{Overall architecture of the {\model} framework.}
    \label{fig:PRS}
\end{figure}

\subsection{Permutation-Matching ({\RM}) stage}
As motivated, the Permutation-Matching ({\RM}) stage is proposed to obtain multiple effective candidate lists in a more efficient way. As shown in the left part of Fig.~\ref{fig:PRS}, various algorithms~\citeN{Ai:DLCM, Pei:PRM, pang2020setrank, Zhuang:MIDNN} can be deployed in parallel to generate item lists, followed by the list merge operation into the candidate list set. Though effective, current efforts neglect the permutation-variant influence within the final item list. To leverage the permutation-variant influence in a efficient way, we propose a permutation-wise and goal-oriented beam search algorithm named {\FLSA} (\textbf{F}ast \textbf{P}ermutation \textbf{S}earching \textbf{A}lgorithm). 

\subsubsection{Offline training.}
To make users take more browsing and interactions, besides the normally applied CTR~\citeN{Zhou:DIN, Feng:DSIN} score, we first design the NEXT score, which predicts the probability whether the user will continue browsing after the item. Items with higher NEXT scores can increase the user's desire for continuous browsing, which improves the probability of subsequent items being browsed and clicked. 

Specifically, with the labeled interaction records $\mathcal{R}$, we elaborate on two point-wise models: the CTR prediction model $M^\text{CTR}(v|u;\mathbf{\Theta}^\text{CTR})$ and the NEXT prediction model 
$M^\text{NEXT}(v|u;\mathbf{\Theta}^\text{NEXT})$, which can be calculated as follows:
\begin{equation} \label{eq:FLSA_eq1}
\begin{aligned}
\hat{y}_{t}^\text{CTR} &= M^\text{CTR}(v|u;\mathbf{\Theta}^\text{CTR}) \\
&= \sigma\bigl(f(f(f(\mathbf{x}_v \oplus \mathbf{x}_u)))\bigr) \\ 
\hat{y}_{t}^\text{NEXT} &= M^\text{NEXT}(v|u;\mathbf{\Theta}^\text{NEXT}) \\
&= \sigma\bigl(f(f(f(\mathbf{x}_v \oplus \mathbf{x}_u)))\bigr)
\end{aligned}
\end{equation}
where $f(\textbf{x}) = ReLU(\textbf{W}\textbf{x} + \textbf{b})$~\footnote{Note that the parameters among different $f(\cdot)$s are not shared.}, $\sigma(\cdot)$ is the logistic function. Clearly, the two models can be optimized via binary cross-entropy loss function, which is defined as follows:
\begin{equation} \label{eq:FLSA_eq2}
    \begin{split}
    \mathcal{L}^\text{CTR} =& -\frac{1}{N}\!\sum_{(u, \mathcal{V}) \in \mathcal{R}}\sum_{x_v^t \in \mathcal{V}} \Big(y_{t}^\text{CTR}\,\log \hat{y}_{t}^\text{CTR} \\
    &+ (1 {-} y_{t}^\text{CTR})\,\log(1 {-} \hat{y}_{t}^\text{CTR})\Big) \\
    \mathcal{L}^\text{NEXT} =& -\frac{1}{N}\!\sum_{(u, \mathcal{V}) \in \mathcal{R}}\sum_{x_v^t \in \mathcal{V}} \Big(y_{t}^\text{NEXT}\,\log \hat{y}_{t}^\text{NEXT} \\
    &+ (1 {-} y_{t}^\text{NEXT})\,\log(1 {-} \hat{y}_{t}^\text{NEXT})\Big)
    \end{split}
\end{equation}
We train $\mathbf{\Theta}^\text{CTR}$ and $\mathbf{\Theta}^\text{NEXT}$ till converged according to Eq.\ref{eq:FLSA_eq2}. 

\subsubsection{Online serving.}
In this work, we regard the re-ranking task as selecting items from the input ranking list sequentially till the pre-defined length. Beam search~\citeN{wiseman2016sequence, levy2007speakers} is a common technology to generate multiple effective lists by exploring and expanding the most promising candidates in a limited set. In the {\FLSA} algorithm, we implement the beam search algorithm in a goal-oriented way, that is, we select the lists with the higher estimated reward at each step.

\begin{figure}
    \centering
    \includegraphics[scale=0.9]{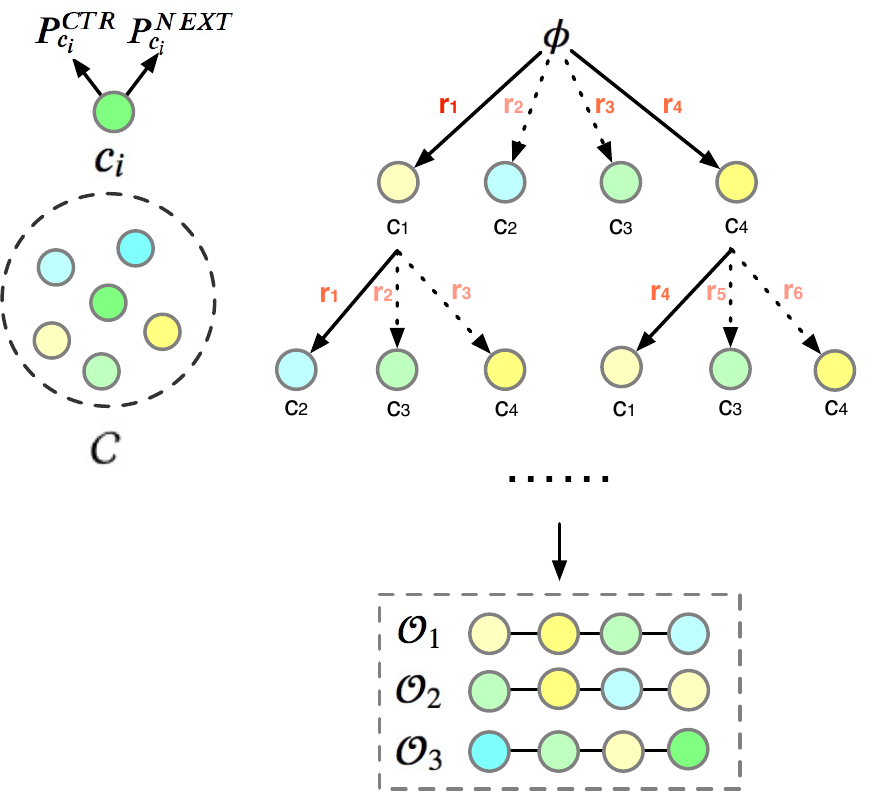}
    \caption{Illustration of the {\FLSA} algorithm.}
    \label{fig:flas_pic}
\end{figure}

We clearly present and illustrate the proposed {\FLSA} algorithm in Fig.~\ref{fig:flas_pic} and Alg.~\ref{alg:FLAS_alg}. First, we attach two predicted scores to each item $c_i$ in the input ranking list $\mathcal{C}$ by the converged CTR prediction model $M^\text{CTR}(v|u;\mathbf{\Theta}^\text{CTR})$ and the NEXT prediction model $M^\text{NEXT}(v|u;\mathbf{\Theta}^\text{NEXT})$: 
the click-through probability $P_{c_i}^\text{CTR}$ and the continue-browsing probability $P_{c_i}^\text{NEXT}$. Afterwards, at each step we exhaustively expand the remaining candidates for each list in the set $\mathcal{S}$ and reserve the top $k$ of candidate lists according to their calculated estimated reward (lines 1-17).
In line 18-28, At each step when iterating through each item in the list, we calculate the transitive expose probability $p^{Expose}$ multiplied by the NEXT scores of previous items. Influenced by $p^{Expose}$, we calculate PV reward $r^{PV}$ and IPV reward $r^{IPV}$, and denote the sum of them as the reward $r^{sum}$ at the end of iteration. Here $r^{PV}$, $r^{IPV}$ and $r^{sum}$ represents the estimated PV, IPV and mixed reward of the list, respectively.

At the end of the algorithm, we obtain the candidate list set $\mathcal{S} = \{\mathcal{O}_1, \dots, \mathcal{O}_t\}_{size=k}$, which can be either directly selected the top list according to the reward $r^{sum}$. Moreover, the final item lists generated from other methods (\eg DLCM and PRM) can be merged into $\mathcal{S}$ and further ranked by the successive {\RR} stage.

\subsubsection{Efficiency Study.} We make some efforts to improve the efficiency of {\FLSA} algorithm. Specifically, we deploy the CTR and NEXT prediction model in parallel in the ranking stage to rate CTR and NEXT scores for each item, whose algorithm complexity is $O(1)$. For the algorithm in Alg.~\ref{alg:FLAS_alg}, we conduct the loops in lines 6-15 in parallel and adopt the min-max heaps sorting algorithm~\cite{atkinson1986min} to select top-k results in line 16. Totally, we reduce to the algorithm complexity if Alg.~\ref{alg:FLAS_alg} to $O(n(n + k \log k))$. 

Overall, the algorithm complexity of {\FLSA} is $a O(1) + b O(n(n + k \log k))$. It is more efficient than existing list-wise methods (\eg DLCM~\cite{Ai:DLCM} and PRM~\cite{Pei:PRM}) of $a O(n) + b O(n \log n)$. Here $a \gg b$ represents that the matrix calculations in deep models cost much more than the numerical calculations.

\begin{algorithm}
\caption{Fast Permutation searching algorithm ({\FLSA}).} 
\label{alg:FLAS_alg} 
\begin{algorithmic}[1] 
\Require 
Input ranking list $\mathcal{C}$; CTR score list $\mathcal{P}^\text{CTR}$; NEXT score list $\mathcal{P}^\text{NEXT}$; Output length $n$; Beam size integer $k$; Fusion coefficient float $\alpha, \beta$.
\Ensure Candidate list set $\mathcal{S}$.
\State New candidate list set $\mathcal{S} = \{[\phi]\}$, estimated reward set $\mathcal{R} = \{\}$;
\State $//$ \textbf{Beam search.}
\For{$i = 1, 2, ..., n$}:
    \State New candidate list set $\mathcal{S}_t = \mathcal{S}$;
    \State Clear $\mathcal{S}$ and $\mathcal{R}$;
    \For{$\mathcal{O} \in \mathcal{S}_t$}:
        \For{$c_i \in \mathcal{C}$}:
            \If{$c_i$ $\notin \mathcal{O}$}:
                \State New $\mathcal{O}_t$ by appending $c_i$ to $\mathcal{O}$ ;
                \State Reward $r$ =  Calculate-Estimated-Reward($\mathcal{O}_t$);
                \State $\mathcal{R} \leftarrow \mathcal{R} \cup \{r\}$;
                \State $\mathcal{S} \leftarrow \mathcal{S} \cup \{\mathcal{O}_t\}$;
            \EndIf
        \EndFor
    \EndFor
    \State According to $\mathcal{R}$, $\mathcal{S} \leftarrow$ top $k$ of $\mathcal{S}$ ;
\EndFor
\State $//$ \textbf{Calculate reward.}
\Function{Calculate-Estimated-Reward}{$\mathcal{O}$}
    \State Estimated PV reward $r^{PV}$ = 1, IPV reward $r^{IPV}$ = 0;
    \State Transitive expose probability $p^{Expose}$ = 1;
    \For{$c_i \in \mathcal{O}$}:
        \State $r^{PV}$ += $P^{Expose} * P^\text{NEXT}_{c_i}$;
        \State $r^{IPV}$ += $P^{Expose} * P^\text{CTR}_{c_i}$;
        \State $p^{Expose}$ $*$= $P^\text{NEXT}_{c_i}$;
    \EndFor
    \State $r^{sum}$ = $\alpha * r^{PV} + \beta * r^{IPV}$;
    \State \Return $r^{sum}$;
\EndFunction
\end{algorithmic}
\end{algorithm}

\subsection{Permutation-rank ({\RR}) stage}
As motivated, the {\RR} stage is proposed to provide a unified permutation-wise ranking criterion for the candidate list set generated by the {\RM} stage. Most methods in current RS mainly follow the greedy strategy based on the rating scores. Though effective, such strategy neglects the permutation-variant influence in the final item list, thus is not capable to evaluate the permutations. To this end, as show in the right part of Fig.~\ref{fig:PRS}, we propose the LR (\textbf{L}ist \textbf{R}eward) metric to select the permutation-optimal list from the candidate list set provided by the {\RM} stage, which is calculated by the rating scores of elaborately designed permutation-wise model {\DCWN} (\textbf{D}eep \textbf{P}ermutation-\textbf{W}ise \textbf{N}etwork).

\begin{figure}
    \centering
    \includegraphics[scale=0.42]{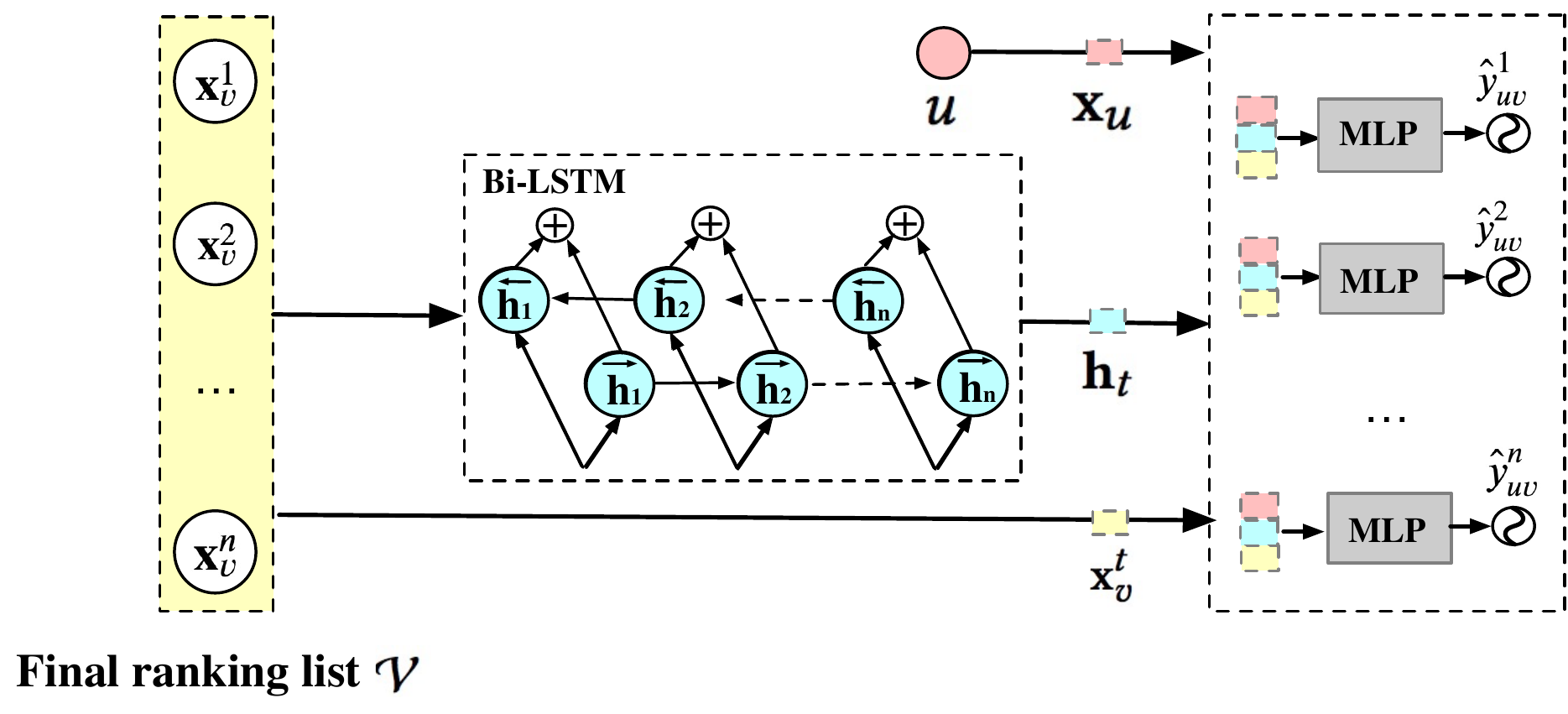}
    \caption{Technical architecture of the {\DCWN} model.}
    \label{fig:Evaluator}
\end{figure}

\subsubsection{Offline training.}
The overall architecture of {\DCWN} is shown in Fig.~\ref{fig:Evaluator}. As motivated, {\DCWN} is proposed to capture and model the permutation-variant influence contained by the final item list. By taking such motivation into consideration, {\DCWN} $M(\mathbf{x}_v^t|u, \mathcal{V}; \mathbf{\Theta}^{D})$, parameterized by $\mathbf{\Theta}^{D}$, predicts the permutation-wise interaction probability between user $u$ and the $t$-th item $\mathbf{x}_v^t$ in the final item list $\mathcal{V}$. 
Naturally, Bi-LSTM~\cite{Graves:bilstm} is excellent at capturing such time-dependent and long-short term information within the permutation. Mathematically, the forward output state for the $t$-th item $x_v^t$ can be calculated as follows:

\begin{equation} \label{eq:eva-bi-lstm}
\begin{aligned}
\textbf{i}_t &= \sigma(\textbf{W}_{xi}\mathbf{x}_v^t + \textbf{W}_{hi}\textbf{h}_{t-1} + \textbf{W}_{ci}\textbf{c}_{t-1} + \textbf{b}_i)\\
\textbf{f}_t &= \sigma(\textbf{W}_{xf}\mathbf{x}_v^t + \textbf{W}_{hf}\textbf{h}_{t-1} + \textbf{W}_{cf}\textbf{c}_{t-1} + \textbf{b}_f)\\
\textbf{c}_t &= \textbf{f}_{t}\mathbf{x}_v^t + \textbf{i}_{t}\tanh(\textbf{W}_{xc}\mathbf{x}_v^t + \textbf{W}_{hc}\textbf{h}_{t-1} + \textbf{b}_{c})\\
\textbf{o}_t &= \sigma(\textbf{W}_{xo}\mathbf{x}_v^t + \textbf{W}_{ho}\textbf{h}_{t-1} + \textbf{W}_{co}\textbf{c}_{t} + \textbf{b}_o)\\
\overrightarrow{\textbf{h}_{t}} &= \textbf{o}_{t}\tanh(\textbf{c}_t)\\
\end{aligned}
\end{equation}
where $\sigma(\cdot)$ is the logistic function, and $\textbf{i}$, $\textbf{f}$, $\textbf{o}$, and $\textbf{c}$ are the input gate, forget gate, output gate and cell vectors which have the same size as $\mathbf{x}_v^t$. Shapes of weight matrices are indicated with the subscripts. Similarly, we can get the backward output state $\overleftarrow{\textbf{h}_{t}}$. Then we concatenate the two output states $\textbf{h}_{t} = \overrightarrow{\textbf{h}_{t}} \oplus \overleftarrow{\textbf{h}_{t}}$ and denote $\textbf{h}_{t}$ as the sequential representation of $x_v^t$.   

Due to the powerful ability in modeling complex interaction in the CTR prediction field~\citeN{Zhou:DIN,Feng:DSIN,Pi:MIMN,zhou2019deep}, we integrate the multi-layer perceptron (MLP) into {\DCWN} for better feature interaction. Hence, we formalize the Bi-LSTM as follows:
\begin{equation} \label{eq:eva-mlp}
    \begin{split}
    M(\mathbf{x}_v^t|u, \mathcal{V}; \mathbf{\Theta}^{D}) = \sigma\bigl(f(f(f(\mathbf{x}_u \oplus \mathbf{x}_v^t \oplus \textbf{h}_{t})))\bigr)
    \end{split}
\end{equation}
where $f(\textbf{x}) = ReLU(\textbf{W}\textbf{x} + \textbf{b})$, $\sigma(\cdot)$ is the logistic function. The parameter set of {\DCWN} is $\mathbf{\Theta}^{D}= \{\mathbf{W}_*, \mathbf{b}_*\}$, \ie the union of parameters for Bi-LSTM and MLP.

Clearly, {\DCWN} can be optimized via binary cross-entropy loss function, which is defined as follows:
\begin{equation} 
    \begin{split}
    \mathcal{L}^{D} = -\frac{1}{N}\!\sum_{(u, \mathcal{V}) \in \mathcal{R}}\sum_{x_v^t \in \mathcal{V}} \left(y_{uv}^t\,\log \hat{y}^t_{uv} + (1 {-} y_{uv}^t)\,\log(1 {-} \hat{y}_{uv}^t)\right)
    \end{split}
\end{equation}
where $\mathbb{D}$ is the training dataset. For convenience, we refer $\hat{y}_{uv}^t$ as $M(\mathbf{x}_v^t|u, \mathcal{V}; \mathbf{\Theta}^{D})$, \ie $\hat{y}_{uv}^t =  M(\mathbf{x}_v^t|u, \mathcal{V}; \mathbf{\Theta}^{D})$, and $y_{uv}^t \in \{0, 1\}$ is the ground truth. We can optimize the parameters $\mathbf{\Theta}^D$ through minimizing $\mathcal{L}^D$.

Note that the major difference which distinguishes our proposed {\DCWN} from current list-wise methods~\citeN{Ai:DLCM, pang2020setrank, Pei:PRM} is that {\DCWN} models the \textit{final item list} rather than the \textit{input ranking list}. Practically, the user is more influenced by the permutation information of the exhibited final item list.

\subsubsection{Online serving.}
As mentioned above, we propose a unified permutation-wise LR metric based on the {\DCWN} model, which is applied to achieve the most  permutation-optimal list solution from the candidate list set generated in the {\RM} stage. Specifically, we calculate the LR (list reward) score for each list $\mathcal{O}_t$ in the candidate list set $\mathcal{S}$ as follows:
\begin{equation}
    \begin{split}
    LR(\mathcal{O}_t) = \sum_{x_o^i \in \mathcal{O}_t}\mbox{M}(x_o^i|u, \mathcal{O}_t)\\
    \end{split}
\end{equation}
Afterwards, the list $\mathcal{P}$ with the highest LR score is finally selected and recommended to the user, with the hope of obtaining the permutation-optimal list and most meeting the user's demands.

Finally, we refer the online serving part of {\RM} and {\RR} stages in the {\model} framework as the learned re-ranking strategy $\pi$ to generate the permutation-optimal list $\mathcal{P}$ from the input ranking list $\mathcal{C}$.

\section{Experiments}
In this section, we perform a series of online and offline experiments to verity the effectiveness of the proposed framework {\model}, with the aims of answering the following questions:
\begin{itemize}
    \item \textbf{RQ1}: How does the proposed LR metric in the {\RR} stage evaluate the profits of the permutations precisely?
    \item \textbf{RQ2}: How does the proposed {\FLSA} algorithm in the {\RM} stage generate the candidate list set from exponential list solutions in the effective way?
    \item \textbf{RQ3}: How about the performance of {\model} in the real-world recommender systems?
\end{itemize}

\begin{table}
\caption{Statistics of datasets}
\label{table:dataset}
\begin{tabular}{lcc}
\toprule
{Description} & {Rec} & {Ad}\\
\midrule
{\#Users} & {5.48 $\times 10^7$} & {1.06 $\times 10^6$} \\
{\#Items} & {2.08 $\times 10^7$} & {8.27 $\times 10^5$} \\
{\#Records} & {5.08 $\times 10^8$} & {2.04 $\times 10^6$} \\
{\#User-item interactions} & {1.44 $\times 10^8$} & {2.04 $\times 10^7$} \\
{\#Avg size of $\mathcal{C}_u$} & {147.69} & {100.0} \\
{\#Avg size of $\mathcal{V}_u$} & {4.0} & {10.0} \\
\bottomrule
\end{tabular}
\end{table}

\subsection{Experimental Setup}
\subsubsection{Datasets}
We conduct extensive experiments on two real-world datasets: a proprietary dataset from Taobao application and a public dataset from Alimama, which are introduced as follows:
\begin{itemize}
    \item \textbf{Rec}~\footnote{www.taobao.com} dataset consists of large-scale list interaction logs collected from Taobao application, one of the most popular e-commerce platform. Besides, Rec dataset contains user profile (\eg id, age and gender), item profile (\eg id, category and brand), the input ranking list provided by the previous stages for each user and the labeled final item list. 
    \item \textbf{Ad}~\footnote{https://tianchi.aliyun.com/dataset/dataDetail?dataId=56} dataset records interactions between users and advertisements and contains user profile (\eg id, age and occupation), item profile (\eg id, campaign and brand). According to the timestamp of the user browsing the advertisement, we transform records of each user and slice them at each 10 items into list records. Moreover, we mix an additional 90 items with the final item list as the input ranking list to make it more suitable for re-ranking.
\end{itemize}
The detailed descriptions of the two datasets are shown in Tab.~\ref{table:dataset}. For the both datasets, we randomly split the entire records into training and test set, \ie we utilize 90\% records to predict the remaining 10\% ones~\footnote{We hold out 10\% training data as the validation set for parameter tuning.}. 
 

\subsubsection{Baselines} 
We select two kinds of representative methods as baselines: point-wise and list-wise methods, which are widely adopted in most industrial recommender systems. Point-wise methods (\ie DNN and DeepFM) mainly predict the interaction probability for the given user-item pair by utilizing raw features derived from user and item profile. List-wise methods (\ie MIDNN, DLCM and PRM) devote to extracting list-wise information in different ways. The comparison methods are given below in detail:
\begin{itemize}
    \item \textbf{DNN}~\cite{Paul:YoutubeNet} is a standard deep learning method in the industrial recommender system, which applies MLP for complex feature interaction. 
    
    \item \textbf{DeepFM}~\cite{Guo:DeepFM} is a general deep model for recommendation, which combines a factorization machine component and a deep neural network component.
    
    \item \textbf{MIDNN}~\cite{Zhuang:MIDNN} extracts list-wise information of the input ranking list with complex handmade features engineering.
    
    \item \textbf{DLCM}~\cite{Ai:DLCM} firstly applies GRU to encode the input ranking list into a local vector, and then combine the global vector and each feature vector to learn a powerful scoring function for list-wise re-ranking.
    
    \item \textbf{PRM}~\cite{Pei:PRM} applies the self-attention mechanism to explicitly capture the mutual influence between items in the input ranking list.
    
    \item \textbf{{\DCWN}} is the elaborately designed permutation-wise model in this work, based on which we propose the LR metric.
\end{itemize}

To answer \textbf{RQ1}, we denote the sum rating scores of the recorded final item list from the baselines ($SR$ for distinction), and our proposed LR metric is calculated by {\DCWN} with the same pattern. 

To answer \textbf{RQ2}, we compare the ranking result of the baselines with the list of highest estimated reward $r^{sum}$ in {\FLSA} under the LR metric. 

To answer \textbf{RQ3}, we deploy the {\RR} stage with {\FLSA} algorithm and the {\RM} stage with the LR metric based on the {\DCWN} model and report the online performance of each stage and the complete {\model} framework.

Note that pair-wise and group-wise methods are not selected as baselines in our experiments due to their high training or inference complexities ($\mathcal{O}(N^2)$) compared with point-wise ($\mathcal{O}(1)$) or list-wise models ($\mathcal{O}(N)$), which may not be appropriate for the industrial RS.

\subsubsection{Evaluation Metrics} 
We apply Loss (cross-entropy) and AUC (area under ROC curve) to evaluate the accuracy of model predictions. Moreover, to answer \textbf{RQ1}, we calculate the pearson correlation coefficient~\cite{benesty:pearson} (Pearson for short) of different metrics \wrt the ground-truth, which refers to the sum of $\mathcal{Y}^\text{CTR}$ (\ie list $IPV$) for each list interaction record in $\mathcal{R}$. Overall, higher Pearson score indicates the estimated list rewards of the model are more correlated to the real ones.

To illustrate the priority of our proposed {\FLSA} algorithm to competitors for \textbf{RQ2}, we adopt the proposed LR metric to the ranking results of the {\FLSA} algorithm and other methods. Besides, we also present the relative improvement~\citeN{feng2020atbrg,feng2020mtbrn} (RI) w.r.t. the LR metric of the {\FLSA} algorithm achieves over the compared methods.

For online A/B testing for \textbf{RQ3}, we choose PV and IPV metrics, which most industrial recommender systems mainly pay attention to. PV
and IPV are defined as the total number of items browsed and clicked by the users, respectively.

\subsubsection{Implementation}
We implement all models in Tensorflow 1.4. For fair comparison, pre-training, batch normalization and regularization are not adopted in our experiments. We employ random uniform to initialize model parameters and adopt Adam as optimizer using a learning rate of 0.001. Moreover, embedding size of each feature is set to 8 and the architecture of MLP is set to [128, 64, 32]. We run each model three times and reported the mean of results. For the {\FLSA} algorithm in Rec dataset, the beam size $k$ is set to $50$ and the output length $n$ is set to $4$.

\subsubsection{Significance Test}
For experimental results in Tab.~3 and 4, we use ``*'' to indicate that the best method is significantly different from the runner-up method based on paired t-tests at the significance level of 0.01.

\begin{table}
\caption{Overall performance comparison \wrt interaction probability prediction (bold: best; underline: runner-up).}
\label{table:model_performance_eva}
\begin{adjustbox}{max width=\linewidth}
\begin{tabular}{lcccccc}
\toprule
\multirowcell{2}{Model} & \multicolumn{3}{{c}}{Rec} & \multicolumn{3}{{c}}{Ad} \\
\cmidrule(lr){2-4} \cmidrule(lr){5-7}
{} & {Loss} & {AUC} & {Pearson} & {Loss} & {AUC} & {Pearson} \\
\midrule 
{DNN($SR$)} & {0.191} & {0.701} & {0.066} & {0.187} & {0.587} & {0.131} \\
{DeepFM($SR$)} & {0.189} & {0.703} & {0.072} & {0.186} & {0.588} & {0.140} \\
\midrule
{MIDNN($SR$)} & {0.182} & {0.706} & {0.090} & {0.185} & {0.600} & {0.155} \\
{DLCM($SR$)} & {0.177} & {0.710} & {0.102} & {0.185} & {0.602} & {0.159}\\
{PRM($SR$)} & {\underline{0.175}} & {\underline{0.712}} & {\underline{0.108}} & {\underline{0.185}} & {\underline{0.602}} & {\underline{0.159}}\\
\midrule
{{\DCWN}(LR)} & {\textbf{0.161}$^\ast$} & {\textbf{0.730}$^\ast$} & {\textbf{0.242}$^\ast$} & {\textbf{0.183}$^\ast$} & {\textbf{0.608}$^\ast$} & {\textbf{0.172}$^\ast$} \\
\bottomrule
\end{tabular}
\end{adjustbox}
\end{table}

\subsection{Performance Comparison (RQ1)}\label{Section_RQ1}
We report the comparison results of {\DCWN} and other baselines on two datasets in Tab.~\ref{table:model_performance_eva}. The major findings from the experimental results are summarized as follows:

\begin{itemize}
    \item Generally, list-wise methods (\ie MIDNN, DLCM and PRM) achieve more improvements in most cases than point-wise methods (\ie DNN and DeepFM). By considering the mutual influence among the input ranking list, these methods learn a refined scoring function aware of the feature distribution of the input ranking list.
    
    \item {\DCWN} consistently yields the best performance on the Loss and AUC metrics of both datasets. In particular, {\DCWN} improves over the best baseline PRM by 0.018, 0.006 in AUC on Rec and Ad dataset, respectively. By taking advantage of the permutation information in the final item list, {\DCWN} shows the ability to provide more precise permutation-wise interaction probabilities of each item. Different from extracting information from the input ranking list in DLCM and PRM, Bi-LSTM helps capture the permutation-variant influence contained by the final item list, which refers as the more essential and effective factors to affect the permutation-wise predictions.
    
    \item It is worth mentioning that $LR$ based on {\DCWN} performs better on the Pearson metric than $SR$ based on other competitors in both datasets. It indicates that {\DCWN}, by leveraging the permutation-variant influence within the final list, can estimate the revenue (\ie IPV) of the permutation more accurately. Hence, its derivative LR metric is accordingly applied to ranking the candidate lists in the {\RR} stage.
    
    \item Compared with the experimental results on the Rec dataset, the performance lift on the Ad dataset is relatively small. One possible reason is that Ad dataset is published with randomly sampling, resulting in the inconsistent and incomplete list records and weaker intra-permutation relevance of user feedback.
\end{itemize}

\subsection{{\FLSA} Algorithm Study (RQ2)}
\begin{table}
\caption{Overall performance comparison \wrt the ranking results on the LR metric (bold: best; underline: runner-up).}
\label{table:model_performance_eva1}
\setlength{\tabcolsep}{1.2mm}{
\begin{tabular}{lcc}
\toprule
Model & LR & RI \\
\midrule 
{DNN} & {0.252} & {+13.09\%} \\
{DeepFM} & {0.261} & {+9.19\%} \\
\midrule
{MIDNN} & {0.272} & {+4.77\%} \\
{DLCM} & {0.274} & {+4.01\%} \\
{PRM} & {\underline{0.278}} & {\underline{+2.51\%}} \\
\midrule
{{\FLSA}$_{\alpha=7, \beta=1}$} & {\textbf{0.285}$^\ast$} & {\textbf{-}} \\
\bottomrule
\end{tabular}}
\end{table}
Firstly, the rationality of using the LR metric to evaluate the model's ranking results offline is listed as follows: 1) As evidenced in the Section~\ref{Section_RQ1}, our proposed LR metric based on {\DCWN} can estimate the list revenue more accurately than the SR metric of other models; 2) It is model-based and independence of the ground-truth, which means can be directly applied to evaluate lists in the {\RR} stage for the online serving and ensure the online and offline consistency.

We report the comparison results of the {\FLSA} algorithm and baselines on the Rec dataset in Tab~\ref{table:model_performance_eva1}~\footnote{Ad dataset is not involved since the sampling of this dataset results in the discontinuity of records.}. We mainly have the following three observations: 
\begin{itemize}
    \item List-wise models (\ie MIDNN, DLCM and PRM) achieve better performance than point-wise models (\ie DNN and DeepFM) on the LR metric, which indicates making the model aware of the feature distribution of the input ranking list helps achieve the better final item list.
    \item Our proposed {\FLSA} algorithm with fine-tuned hyper parameters $\alpha=7, \beta=1$ yields the best performance on the LR metric. In particular, {\FLSA} relatively improves over the strongest competitor PRM by 2.51\%. Different from the normally applied greedy re-ranking strategy, {\FLSA} considers the probability of continuous browsing after the item. By calculating the estimated reward at the each step of beam-searching, {\FLSA} can dynamically balance the demands of user's browsing and interaction, so as to measure the value of the final item list more accurately.
\end{itemize}

\subsubsection{Parameter Sensitivity.} Practically, we have noticed the pre-defined hyper parameters $\alpha$ and $\beta$ have a great influence on the final item list and the LR metric. Therefore, we conduct several grid-search experiments on these two hyper parameters to get a comprehensive understanding of the {\FLSA} algorithm. Specifically, we fix $\beta=1$ and tune $\alpha$ from $0$ to $MAX$, and observe how LR varies under each hyper parameter. Note that setting $\alpha$ to $0$ or $MAX$ means that the final item list is in a descending order \wrt the NEXT or CTR score, respectively.

\begin{figure}
    \centering
    \includegraphics[scale=0.3]{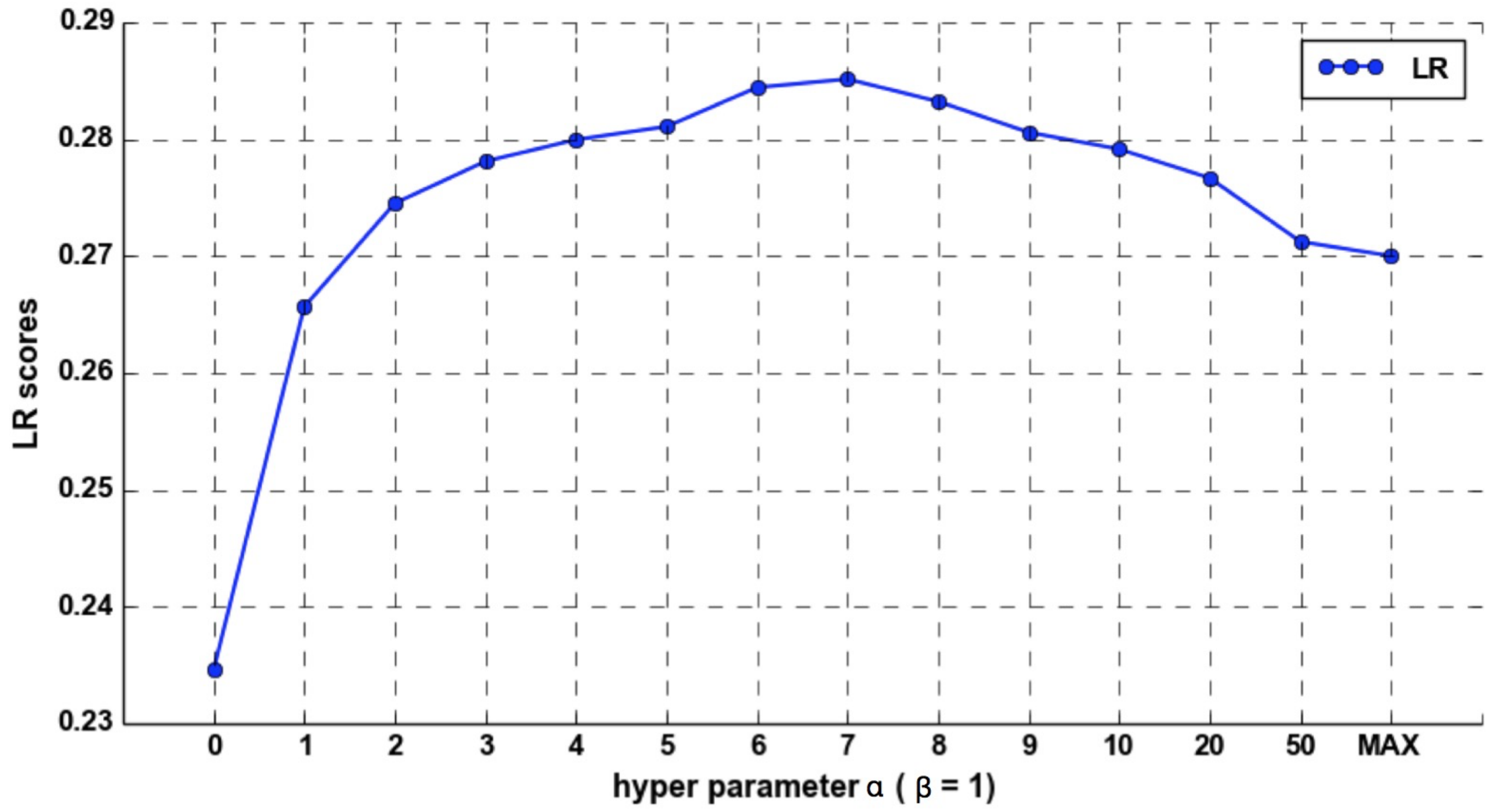}
    \caption{Impact of hyper parameters $\alpha$ on the LR metric.}
    \label{fig:flsa_parameter_sentivity}
\end{figure}

As shown in Fig.~\ref{fig:flsa_parameter_sentivity}, we present the influence of hyper parameters on the final item lists. We observe the LR metric \wrt $\alpha$ declines from $7$ to $0$ and from $7$ to $MAX$, sharply at $0$. It indicates that excessively guiding users to browse or interact may miss his/her demands or increase his/her fatigue, respectively. When setting $\alpha$ to $7$, recommendation results of {\FLSA} take a browsing and interacting balance and deliver the best user experience.

\subsection{Online A/B Testing (RQ3)}

\begin{figure}
    \centering
    \includegraphics[scale=0.06]{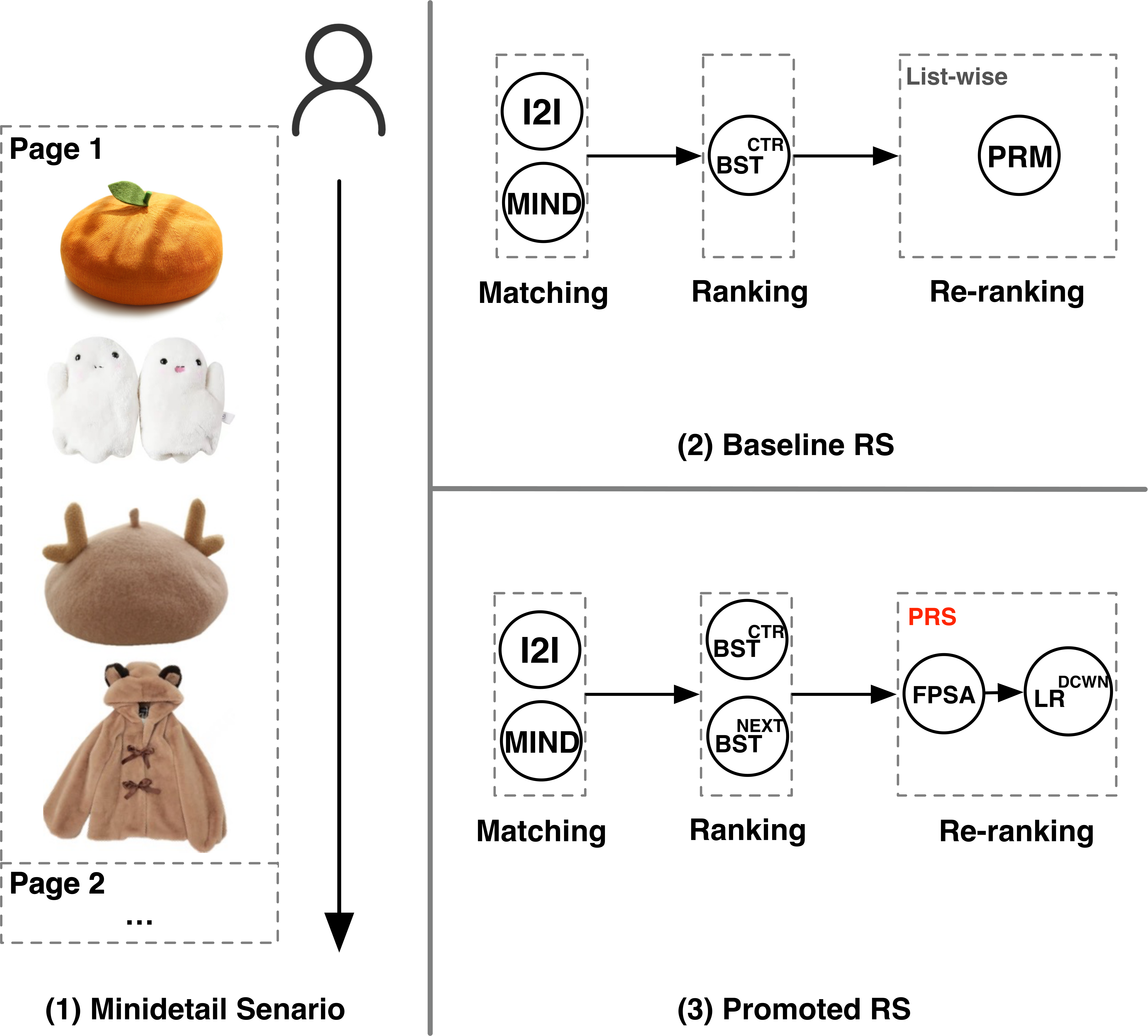}
    \caption{(1) is the \emph{Minidetail} scenario. (2) and (3) present the difference of the baseline RS and the promoted RS with the {\model} framework.}
    \label{fig:online}
\end{figure}

\begin{figure}
    \centering
    \includegraphics[scale=0.14]{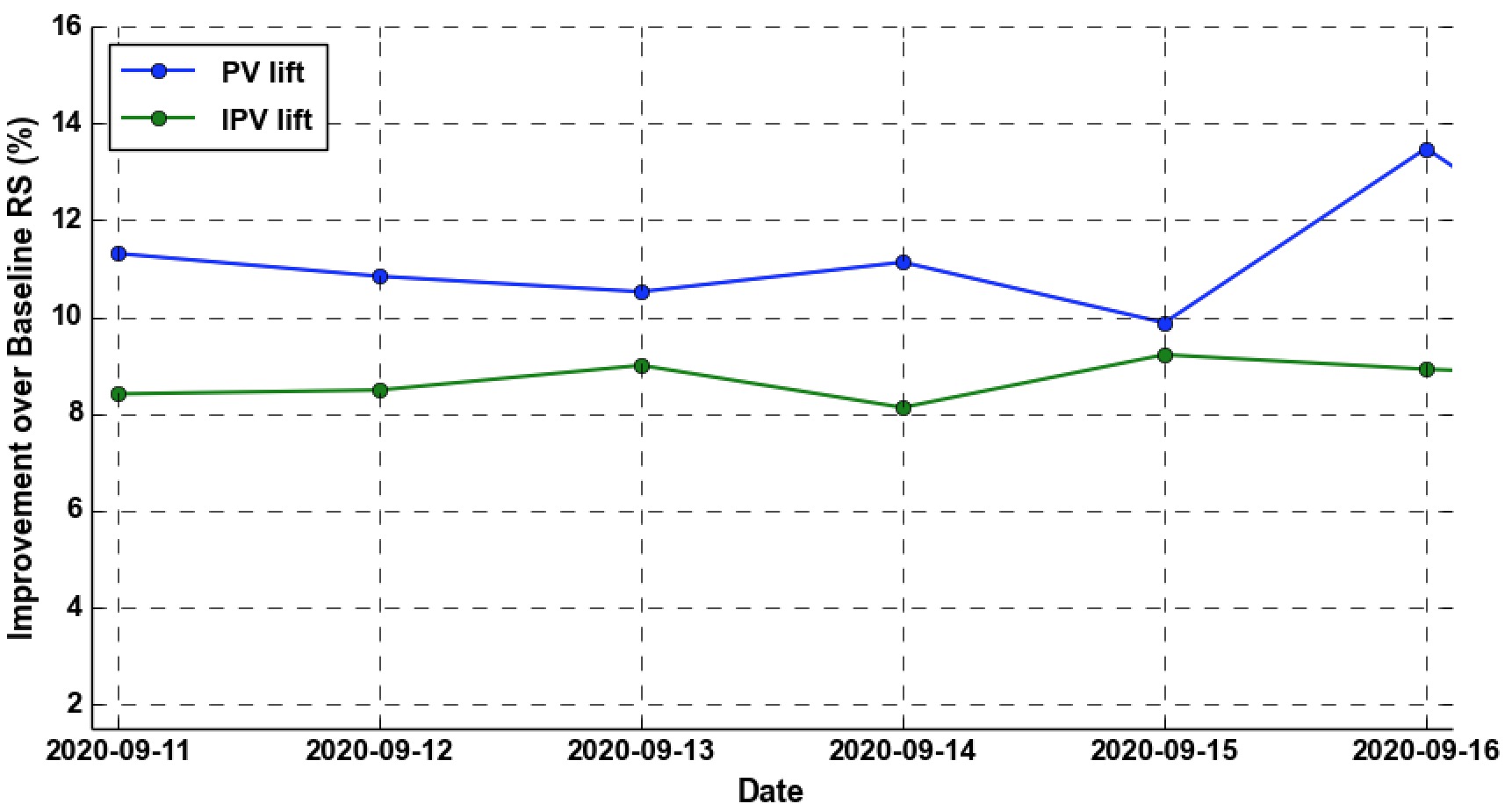}
    \caption{Online improvements on PV and IPV metrics in a week.}
    \label{fig:online_performance}
\end{figure}

To verify the effectiveness of our proposed framework {\model} in the real-world settings, {\model} has been fully deployed in the \emph{Minidetail} scenario, one of the most popular recommendation scenario of Taobao application. AS shown in the left part of Fig.~\ref{fig:online}, in this waterfall scenario, users can browse and interact with items sequentially and endlessly. The fixed-length recommended results are displayed to the user when he/she reaches the bottom of the current page. Considering the potential problem of high latency to the user experience, it brings great challenges to deploy {\model} into production.

We first introduce our effective baseline RS, and then the efforts of merging the {\model} framework into the promoted RS. AS shown in the right part of Fig.~\ref{fig:online}, the baseline RS consists of the matching, ranking and re-ranking stages, equipped with I2I~\cite{i2i} and MIND~\cite{mind} in matching, BST~\cite{bst} in ranking and PRM~\cite{Pei:PRM} in re-ranking. For the {\FLSA} algorithm in the {\RM} stage, we 
concurrently deploy $BST^\text{CTR}$ and $BST^\text{NEXT}$ to calculate the CTR and NEXT scores for each item in the ranking stage, without extra time cost. Then multiple candidate final item lists are generated by the {\FLSA} algorithm, and then ranked by the LR metric calculated by the deployed {\DCWN} model in the {\RR} stage. Finally, the final item list with highest LR scores are recommend to the user.

Thanks to the above efforts, we successfully deployed {\model} into production and report the online relative improvements from "2020-09-11" to "2020-09-17" in the Fig.~\ref{fig:online_performance}. Specifically, the {\FLSA} algorithm in the {\RM} stage obtains \textbf{11.0}\% lift on PV metric and \textbf{2.6}\% lift on IPV metric, with the average cost of \textbf{1.2} milliseconds for online inference. Afterwards, the LR metric based on the {\DCWN} model in the {\RR} stage obtains \textbf{6.1}\% lift on IPV metric, with the average cost of \textbf{6.2} milliseconds for online inference. Totally, the {\model} framework takes only \textbf{7.3} milliseconds to obtain the significant \textbf{11.0}\% lift on PV metric and \textbf{8.7}\% improvement on IPV metric. Considering the massive influenced users and maturity of the baseline RS, the promotion of recommendation performance verifies the effectiveness of our proposed framework {\model}.

\section{Conclusion}
In this paper, we highlight the importance of the permutation knowledge in the final ranking list and address how to leverage and transform such information for better recommendation. We propose a novel two-stage permutation-wise framework {\model} in the re-ranking stage, consisting of the {\RM} and {\RR} stages successively. The {\RM} stage with the proposed {\FLSA} is designed to generate multiple candidate item lists, and the {\RR} stage with the LR metric based on the {\DCWN} model provide a uniform ranking criterion to select the most effective list for the user. Extensive experiments on both industrial and benchmark datasets demonstrate the effectiveness of our framework compared to state-of-the-art point-wise and list-wise methods. Moreover, {\model} has also achieved impressive improvements on the PV and IPV metrics in online experiments after successful deployment in one popular recommendation scenario of Taobao application. 

In the future, we will concentrate on proposing more effective methods in the {\RM} and {\RR} stages. Besides, with the improvement of computing efficiency, {\model} has the potential to replace matching and ranking stages with {\RM} and {\RR} stages, respectively. 


\bibliographystyle{ACM-Reference-Format}
\balance
\bibliography{references}

\end{document}